Rb-NMR study of the quasi-one dimensional competing spin-chain compound

$Rb_2Cu_2Mo_3O_{12}$

Kazuki Matsui, Ayato Yagi, Yukihiro Hoshino, Sochiro Atarashi,

Masashi Hase, Takahiko Sasaki and Takayuki Goto

Rb-NMR study has been performed on the quasi-one dimensional competing spin chain $Rb_2Cu_2Mo_3O_{12}$ with ferromagnetic and antiferromagnetic exchange interactions on nearest neighboring and next nearest neighboring spins, respectively. The system changes from a gapped ground state at zero field to the gapless state at $H_C \simeq 2$ T, where the existence of magnetic order below 1 K was demonstrated by a broadening of NMR spectrum, associated with a critical divergence of $1/T_1$. In higher temperature region, $T_1^{-1}$ showed a power-law type temperature dependence, from which the field dependence of Luttinger parameter $K$ was obtained and compared with theoretical calculations based on the spin nematic Tomonaga Luttinger Liquid (TLL) state.

Despite the simplicity in its Hamiltonian, the one dimensional system still gives us rich and non-trivial physics, such as Tomonga-Luttinger liquid (TLL) and nematic spin state. The title compound $Rb_2Cu_2Mo_3O_{12}$, the quasi-one dimensional quantum spin chain system, involves both of these two phenomena [1-3]. It consists of so-called ribbon-chains of $S = ½$ spins, in which the ferromagnetic and antiferromagnetic exchange interactions work on the nearest neighboring and the next nearest neighboring spins, respectively. This exchange-path configuration, described by $J_1 - J_2$ model, possesses a strong frustration effect and has so far been investigated intensively with expectation to find an exotic ground state. Particularly, in this model, the neighboring two spins tend to form an $S = 1$ spin, being capable of showing the nematic state [4-12]. In the nematic TLL state, while the nemaic operater $S_j^{\pm} S_{j+1}^{\mp}$ and the longitudinal spin $S_j^z$ exibit quasi-long range orders, while the transverse spin correlator $\langle S_j^{\pm} S_0^{\mp} \rangle$ decays exponentially due to the formation of two-magnon bound states [5,15]. In the high-field regime near the saturation, the nematic correlation is stronger than the longitudinal spin correlation, while the latter grows stronger in the low-field region [5,15]. Though the possibility of nematic order or nematic TLL state has so far been studied intensively, there seems to be no consensus on their experimental evidence. One of the difficulties in studying the nematic state lies in the fact that it requires probing four spin-correlation functions with high accuracy [5,13-15].

In order to overcome this difficulty, Sato *et al.* proposed a unique procedure with NMR technique to detect the nematic TLL state [13,14]. The first motivation of this paper is to show an evidence for the nematic TLL state in $Rb_2Cu_2Mo_3O_{12}$ by $^{87}$Rb-NMR combined with Sato's method.

The compound $Rb_2Cu_2Mo_3O_{12}$ was first introduced by one of our colleagues Hase *et al.* as a competing spin chain with dominant Heisenberg interactions of ferromagnetic $J_1/k_\mathrm{B} = -138$ K and antiferromagnetic $J_2/k_\mathrm{B} = 51$ K ($J_1/J_2 = -2.7$) [1-2]. He had also showed that there is no magnetic anomaly at low temperatures down to 2 K. With the large exchange couplings and also the absence of magnetic order at least above 2 K, one expects that the TLL or nematic-TLL state may be realized within a wide temperature range. Recently, Yasui *et al.* has shown that it has a very small spin excitation gap of approximately 2.3 K at zero field. They explain that this gap can roughly be understood in terms of Haldane picture with effective $S = 1$ spin formed by two adjacent spins. The gap is smeared out by a weak magnetic field around $H_\mathrm{C} \simeq 2$ T, and opens up again at high field of around $H_\mathrm{S} \simeq 12$ T [1,2,16,17]. We have recently investigated these two gapped regions by NMR and revealed that the adjacent two spins are coherently excited in the latter field region [18,19]. In this paper, we focus on the intermediate field region, where the system is gapless, to investigate TLL and the field induced magnetic order. Another motivation of this study on this point is to find a crossover point between TLL and paramagnetic state. Usually, the change between the paramagnetic and TLL state is considered to be a crossover, where the system is expected to show only a gradual change rather than a specific temperature point. However, recent many experimental reports on various low dimensional quantum spin systems found rather a clear line between the TLL and the paramagnetic phase [20-22]. This line is conspicuous only in the vicinity of the quantum critical point (QCP), where the system changes from gapped to gapless state or vici versa [20-22]. We will try to find whether or not the line can be seen by NMR in the present system.

Before showing the NMR results, we briefly describe here the experimental details and also procedures for the data analysis. $^{87}$Rb-NMR ($I = \frac{3}{2}$, $\gamma = 13.928$) measurements were performed on the powder sample in the field region from 2 to 12 T, and in the temperature range from 0.3 to 20 K. We will combine our previous data above 12 T and below 2 T, which were already published [18]. The NMR spectra were obtained by plotting the spin-echo amplitude against the applied field. The nuclear spin-lattice relaxation rate was obtained by tracing the spin-echo amplitude against the repetition rate of measurement [23-28].

The present system contains the three Rb sites 4e, 4d and 8f in the unit cell, and the distance between each of these sites and its nearest Cu site is nearly alike distributing between 4.01 and 5.57 Å [1,2]. In the powder spectrum of NMR, these three sites forms a single peak for the central transition between $I_z = \pm½$, combining all of the Knight shift anisotropy, its site difference, and eqq-broadening. This coalition was observed also in the nuclear spin relaxation, which showed a typical single-component relaxation for $I = \frac{3}{2}$ nuclear spin. In Fig. 1, a typical Rb-NMR spectrum and a relaxation curve are shown with the schematic drawing of exchange paths in the ribbon chain.

The hyperfine coupling constant of the Rb site was obtained to be $A = -0.042\mathrm{T}/\mu_\mathrm{B}$ by the scaling between the temperature dependence of macroscopic susceptibility $\chi$ and that of Knight shift [18]. The temperature dependence of the line width (FWHM) was also scaled with $\chi$ to obtain the anisotropic part of the hyperfine coupling as $3A_\mathrm{an} = 0.045\mathrm{T}/\mu_\mathrm{B}$. These assure that both the isotropic and anisotropic part of hyperfine coupling tensor are comparable, the fact of which is crucial in the analysis of $1/T_1$ as will be stated below. We note here that those obtained values are the effective value averaged over the three Rb sites.

Next, we describe the Sato's trick to detect the nematic TLL state. First, in the ordinary (or one-magnon) TLL state, the spatial spin correlation obeys the power law as $\langle S_z(x)S_z(0)\rangle \propto x^{-2K}$ or $\langle S_+(x)S_-(0)\rangle \propto x^{-1/2K}$, where $K$ is the Luttinger parameter, characterizing the TLL state with another parameter of the magnon velocity [9,29]. $K$ shows a characteristic field dependence, which also depends on Hamiltonian. For example, for the Heisenberg antiferromagnetic spin chain, with increasing applied field from zero, $K$ increases monotonically from ½ to 1 at the saturation field, while in gapped systems such as spin ladders, it starts from 1 at the QCP, where the gap is collapsed [29,30]. So far, the field-dependence of this parameter has been studied theoretically for many types of spin chains, including alternating or competing chains [30-37]. The third motivation of our study is to compare the obtained NMR data with those reported theories. Luckily, $K$ can easily be evaluated experimentally by the temperature dependence of NMR-$1/T_1$ as $1/T_1 \propto A_{\parallel}^2 T^{2K-1} + A_{\perp}^2 T^{1/2K-1}$, where $A_{\parallel}$ and $A_{\perp}$ are the hyperfine coupling tensor components, which mediate the longitudinal and transverse spin fluctuation with the nuclear spin relaxation, respectively. Note that according to this formula, $1/T_1$ always diverges at low temperatures irrespective of $K$, except for ½. Contrary to this power-law-type behavior in the ordinary TLL state, $1/T_1$ for the nematic (or two-magnon) TLL state shows a qualitatively different dependence on the applied field. That is, in the nematic TLL state, the transverse component of spin correlation is strongly suppressed and shows an exponential decay [13-14] to give $1/T_1 \propto A_{zz}^2 T^{2K-1}$, which decreases at low temperatures when $K > ½$. This tells us that one can distinguish between the nematic and one-magnon TLL directly by investigating the temperature dependence of $1/T_1$ in the high field region, where the Luttinger parameter $K$ is expected to take the value $K > ½$.

In Fig. 2, we show the temperature dependence of $1/T_1$ measured under various magnetic fields. The upper and lower panels show the data below and above 8.5 T, respectively. For each all data, the temperature dependence of $1/T_1$ obeys the power law in a finite region of temperature. At high temperatures, $1/T_1$ deviates from the power law and tends to stay constant. In Fig. 2, downward arrows indicate this deviation point, denoted as $T_{\mathrm{TL}}$. This change in the temperature dependence of $1/T_1$ may correspond to the cross-over between the paramagnetic state, that is, $T_1 =$ *const.* and the TLL state, that is, the power law of temperature. As increasing the field from $H_{\mathrm{C}}$ or decreasing it from $H_{\mathrm{S}}$, $T_{\mathrm{TL}}$ rapidly increases and exceeds the measured temperature range. Next, at lowest temperatures, $1/T_1$ deviates from the power law again and tends to diverge at lowest temperatures. This deviation, shown by upward arrows in Fig. 2, was observed in the field region between 7 and 9 T, and will be discussed later.

Tracing the power law index of $1/T_1$ with increasing the applied field, one notes that it changes sign from negative to positive at the field 11.1 T. That is, at high field region above, 11.1 T, $1/T_1$ decreases at low temperatures. This indicates that the $T^{1/2K-1}$ term, which corresponds to the transverse component of spin correlation is actually suppressed and does not contributes to $1/T_1$, and hence leads immediately one to the conclusion that the system is in the nematic TLL state at the higher field region [13,14]. With this conclusion, one can proceed and obtain the field dependence of $K$, which is shown in Fig. 3. Note that in order to compare with theories, the abscissa is converted to the uniform magnetization, where the saturation value is defined as 0.5 [13,14]. With increasing field from zero, $K$ decreases first and takes minimum of 0.25 and then increases, heading toward unity in the high field region. We compare this field (or magnetization) dependence of $K$ with the theoretical report of Hikihara *et al*. to find a good agreement for $J_1/J_2 = -2.5$ or $-2.6$ in the higher field region [5]. Furthermore, this $J_1/J_2$ coincides with the value of $-2.6$, which was independently estimated from magnetic measurements on the present compound [1,2].

In the lower field region near $H_C$, the experimentally-determined $K$ showed an upturn with decreasing field. This behavior is consistent with the fact that the present system has a finite spin excitation gap at zero field [16-18] and that $K$ should be unity at QCP where the spin gap is collapsed [9,29]. Note that $K$ reaches ½ at zero field for the gapless Heisenberg chain.

Next, we show in Fig. 4 the spectra taken at the field of 10 T, which is in between $H_C$ and $H_S$. One can see a significant broadening at low temperatures below 1K. The temperature dependence of width (FWHM) was evaluated and also plotted. This steep increase is considered to be due to the emergence of static hyperfine field, and hence the evidence for long-range magnetic order at $T_N \simeq$ 1 K, the sign of which has already been seen as the critical divergence in $1/T_1$ at low temperatures, which are shown by upward arrows in Fig. 2, as described above. These two observations indicate the existence of field-induced magnetic order in the gapless field region. Within the critical region near $T_N$, the temperature dependence of width obeys the mean field theory, that is, $\beta =$ ½, indicating the three dimensional character of this ordering.

The approximate size of hyperfine field due to the magnetic order at 0.3 K is 0.036(2) T. If one assumes the two-sublattice antiferromagnetic (AF) structure or incommensurate SDW, which gives identical flat-top shaped spectra and hence cannot be distinguished by NMR measurements on powder sample, the ordered moment under the field region between 7 and 10 T is roughly estimated to be 0.40(2) $\mu_B$, the value of which is reasonable for the divalent Cu, if one takes into account the spin shrink due to quantum fluctuation. Under the slightly lower field of 5.5 T, the size is reduced to be 0.10(2) $\mu_B$. This result indicates that the high-field ground state is SDW or AF rather than the nematic order. The competing behavior between SDW (or AF) and the nematic order has so far been discussed theoretically [15,41], and the importance of interchain interaction is pointed out. In order to determine the spin structure and also existence of a possible nematic ordered state, measurements on a single crystal or a uniaxially aligned powder sample is indispensable, which is now in the progress.

Finally we show an HT phase diagram in Fig. 5, where we plot against the field the Néel temperature $T_N$ determined from the temperature dependence of NMR line width, the cross-over temperature $T_{TL}$, and also the spin excitation gap, taken from our previous report by Yagi *et al.* [18]. One notes that $T_N$ takes the maximum of 1 K, at the midst of gapless field region between $H_C$ and $H_S$, that is, at around 8 T. This behavior, that is, the bell-shaped dependence of $T_N$ against the applied magnetic field is quite alike as the field-induced magnetic order observed in other quantum spin systems [20-22,38-39].

The overview of Fig. 5 tells us that with increasing the applied field from zero, the spin excitation gap is reduced and is collapsed at around $H_{\mathrm{C}}$ =2 T, then, in the gapless field region, the nematic TLL state appears above 1 K, and the Neel order takes place in succession below 1 K, and at still higher field $H_{\mathrm{S}}$ =12 T, again opens the gap, which increases linearly with the field as $g\mu_{\mathrm{B}}(H - H_{\mathrm{S}})$, where $g \simeq 4$ [18,19]. The crossover temperature $T_{\mathrm{TL}}$ between the paramagnetic state and the nematic TLL state was determined in the vicinity of QCP's, $H_{\mathrm{C}}$ and $H_{\mathrm{S}}$. The field dependence of $T_{\mathrm{TL}}$, that is, steep increase from zero as departing QCP, is quite alike to those reported recently for other spin gap systems [19-21] and also the gapless spin chain[40]. One cannot determine the implication of this apparent boundary at this stage, and further investigation and accumulation of data seem to be important.

In summary, we have investigated the quasi-one-dimensional competing spin chain $Rb_2Cu_2Mo_3O_{12}$ by NMR in the wide field region up to 18 T to find that the system becomes gapless in the field region between $H_{\mathrm{C}}$ =2 T and $H_{\mathrm{S}}$ =12 T, where the system shows the field-induced magnetic order below $T_{\mathrm{N}}$ = 1 K under 8 T. And in the limited temperature region above $T_{\mathrm{N}}$, existence of the nematic TLL state was demonstrated. The field-dependence of the Luttinger parameter was successfully determined from $1/T_1$ and accorded with the theoretical estimation by Hikihara *et al*.

Acknowledgements

This word was supported by JSPS KAKENHI Grant Numbers JP24540350 and JP15K05148. A Part of the experiment was performed at IMR, Tohoku University. The authors are grateful for a kind and valuable discussions with Prof. Masahiro Sato, Prof. Toshiya Hikihara, Prof. Hidekazu Tanaka and Prof. Tomi Ohtsuki.

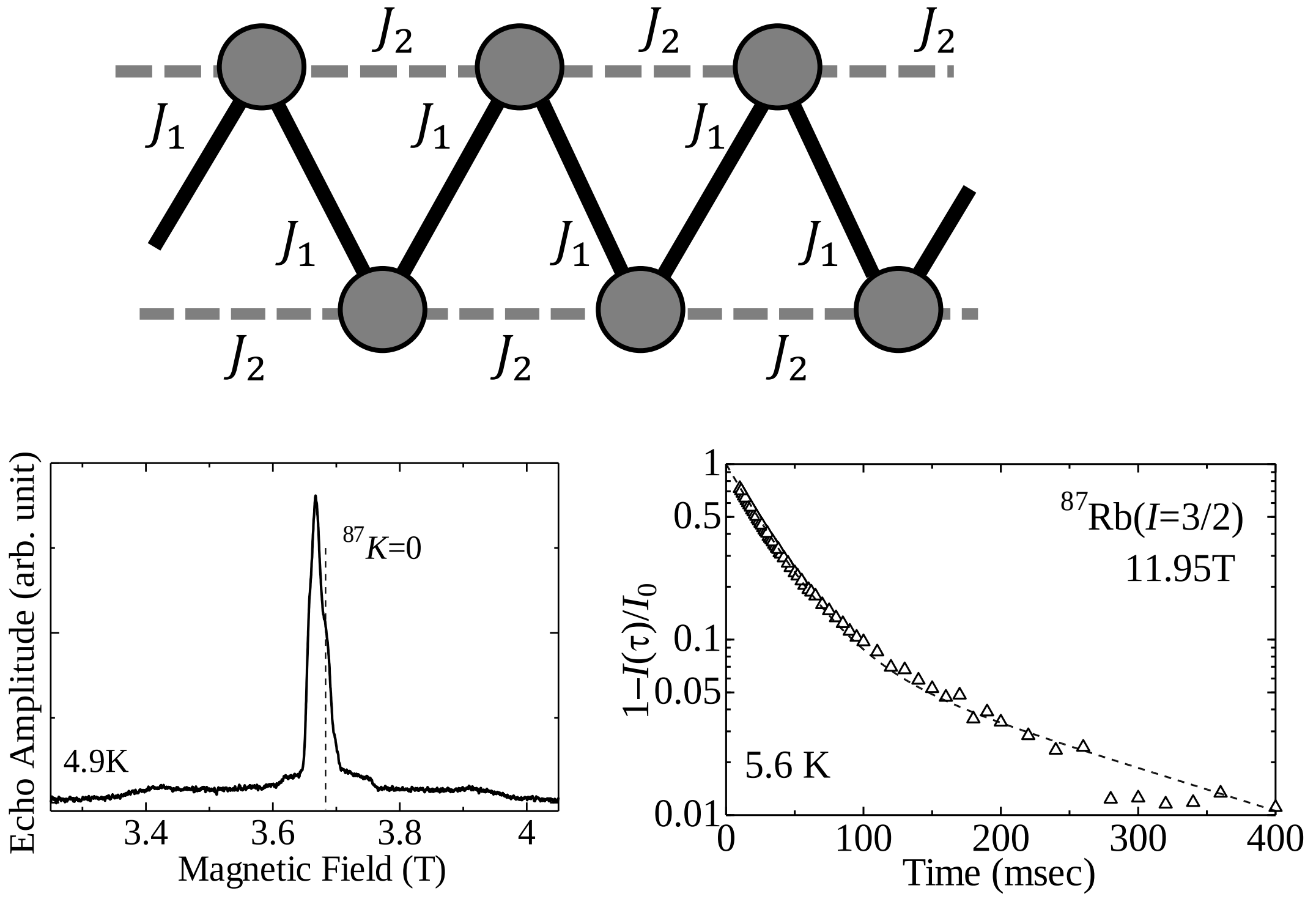

Figure 1. (upper) The schematics of exchange paths on divalent coppers in the ribbon-chain with two dominant interactions of ferromagnetic $J_1$ and antiferromagnetic $J_2$. (lower) A typical $^{87}$Rb-NMR spectrum on the powder sample, with the zero shift position shown by a dashed line, and a typical relaxation curve for $T_1$ measurements, with the fitted relaxation function $0.9e^{-6t/T_1} + 0.1e^{-t/T_1}$ shown by dashed curve.

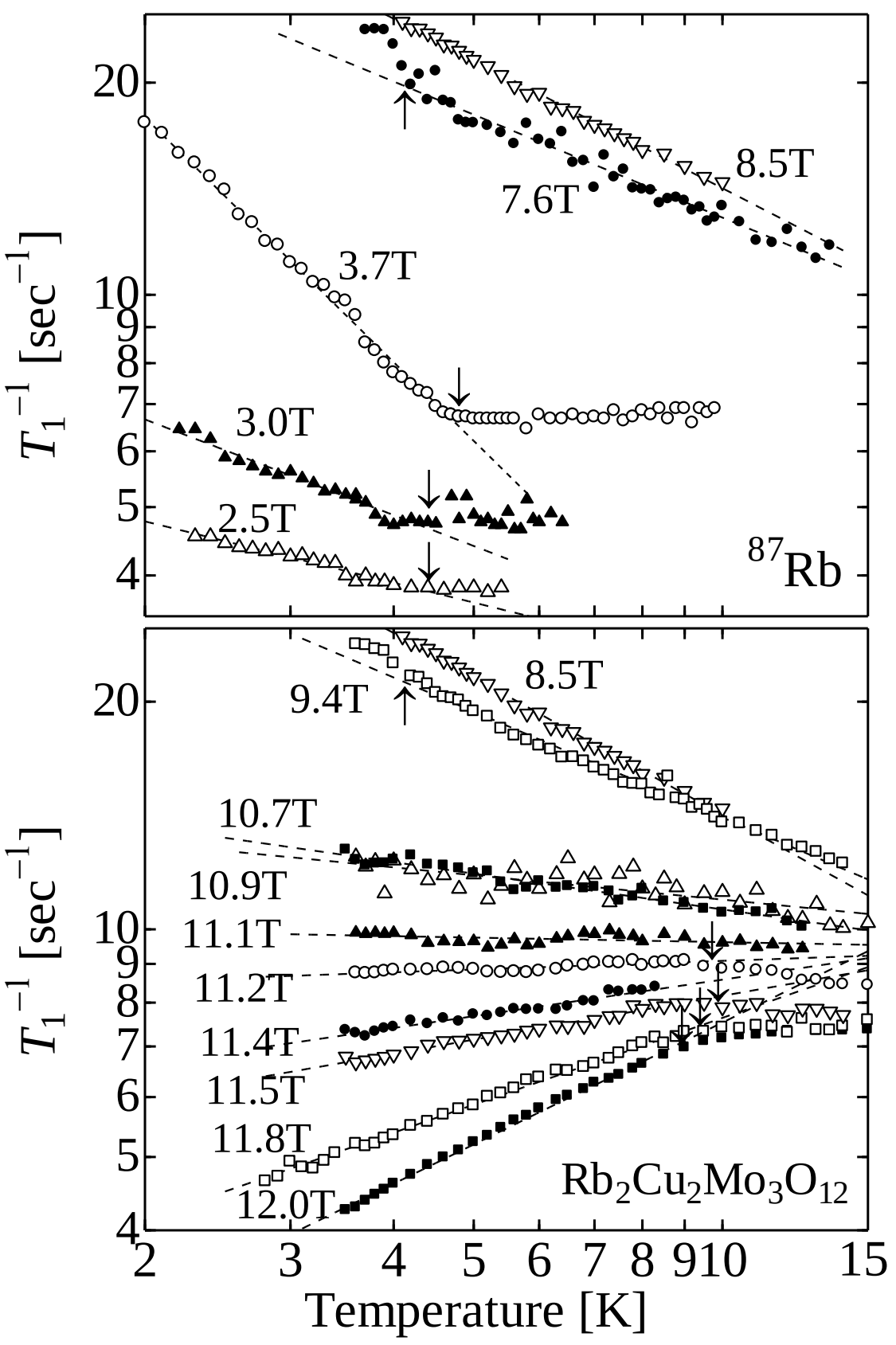

**Figure 2** Temperature dependence of $T_1^{-1}$ under various magnetic field between $H_C \simeq$ 2 and $H_S \simeq$ 12 T. The upper panel shows the data below 8.5 T, and the lower, above. Dashed lines show the fitted function of power law temperature dependence. Downward (upward) arrows show the temperature, where $T_1^{-1}$ deviates from the power law at high (low) temperatures.

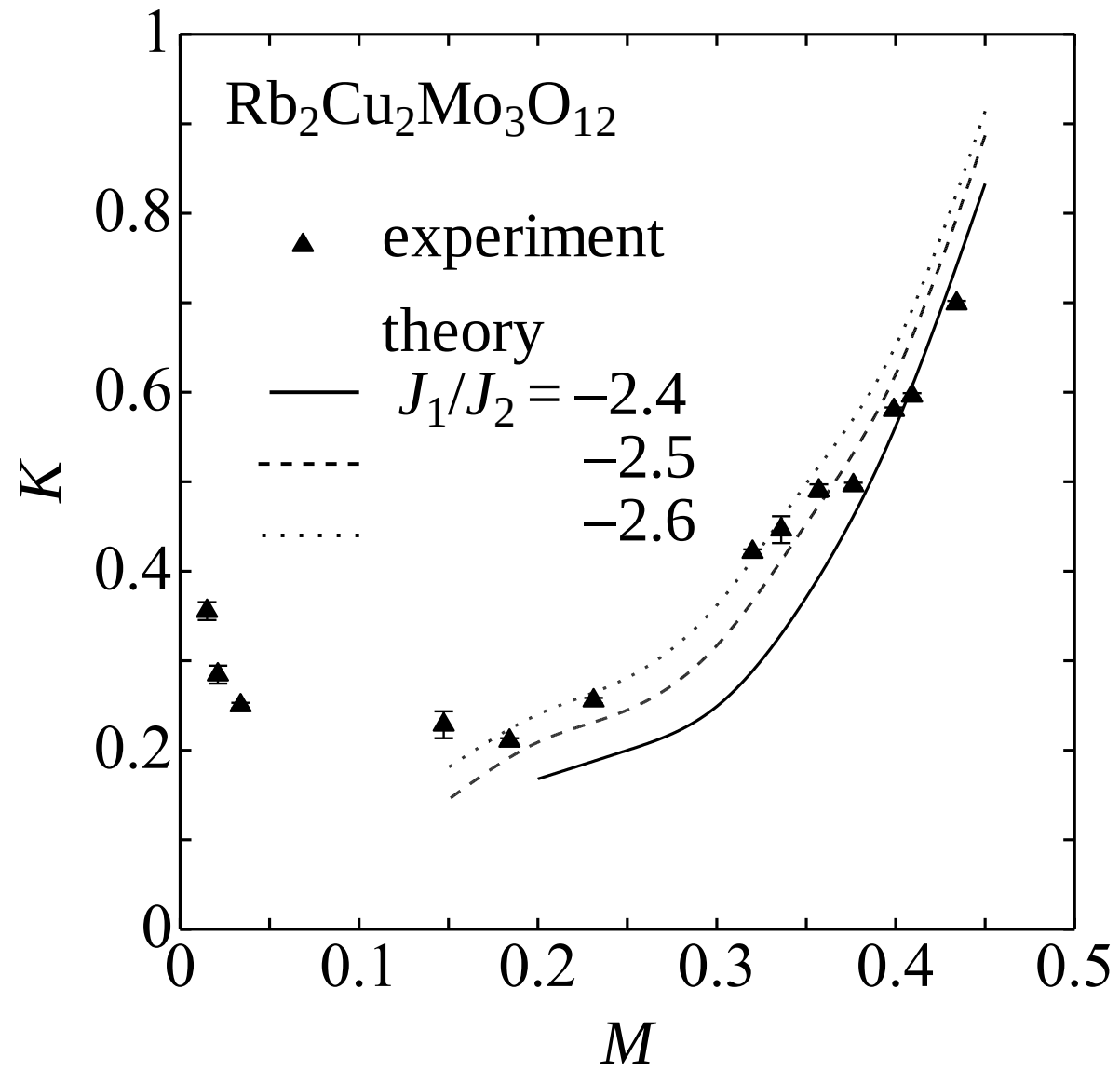

Figure 3 The magnetization ($M$) dependence of the Luttinger parameter $K$, obtained experimentally (open symbols), and that of calculated (curves) by Hikihara *et al.* [35]

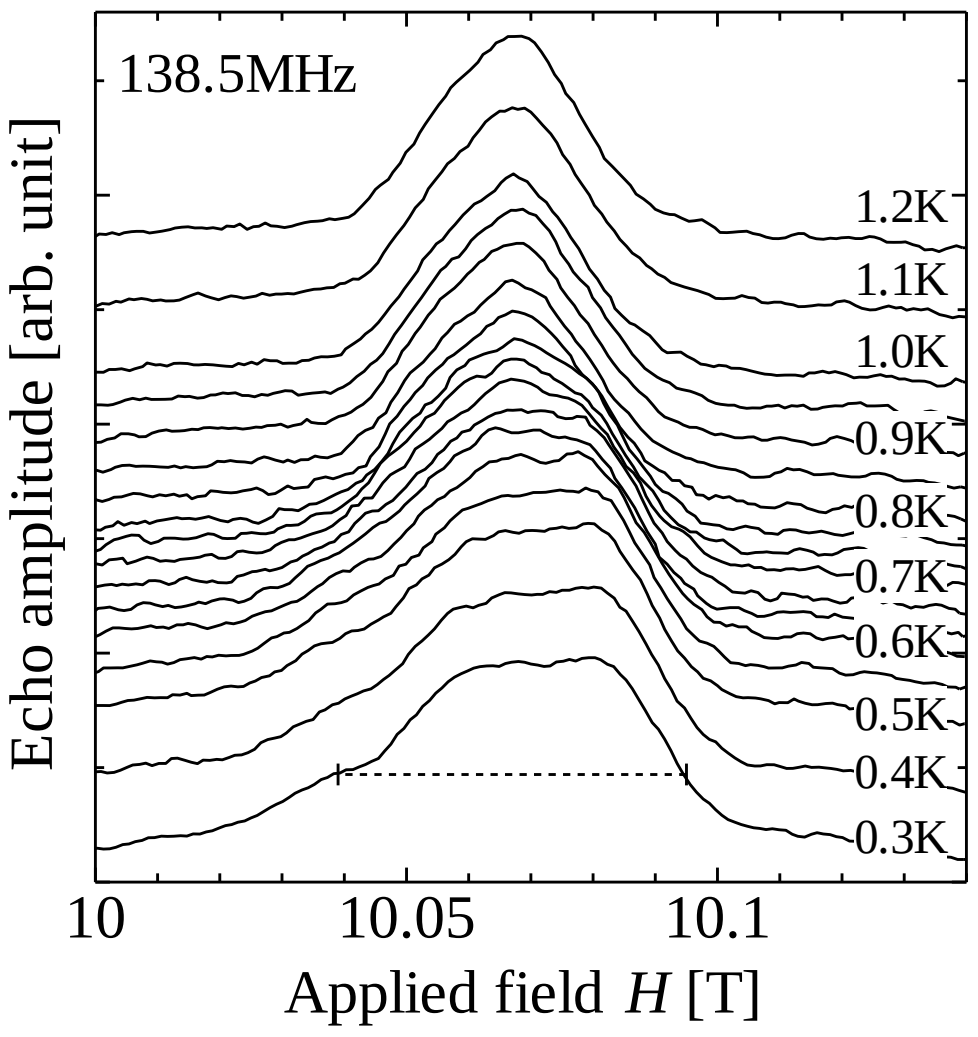

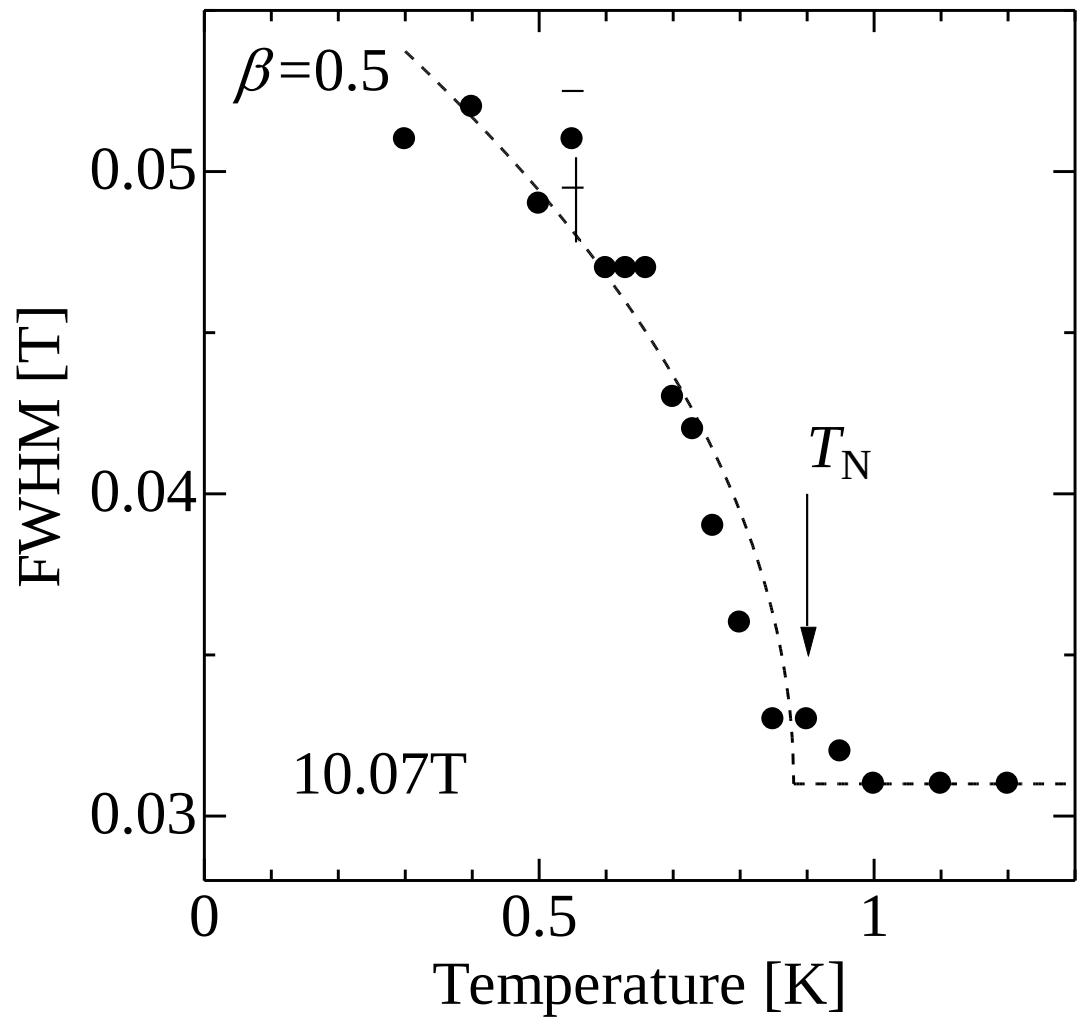

Figure 4 (left) Typical profile of the central peak for $^{87}$Rb-NMR spectra at low temperatures down to 0.3 K. The horizontal dashed line shows the definition of the line width (FWHM), and the vertical dashed line shows the zero shift position. (right) The temperature dependence of FWHM defined in the left panel. The dashed curve shows the critical behavior with $\beta = 0.5$. $T_N$ ($\simeq$ 0.98 K at 10.1 T) shown by an arrow was defined as the onset of the steep increase in FWHM.

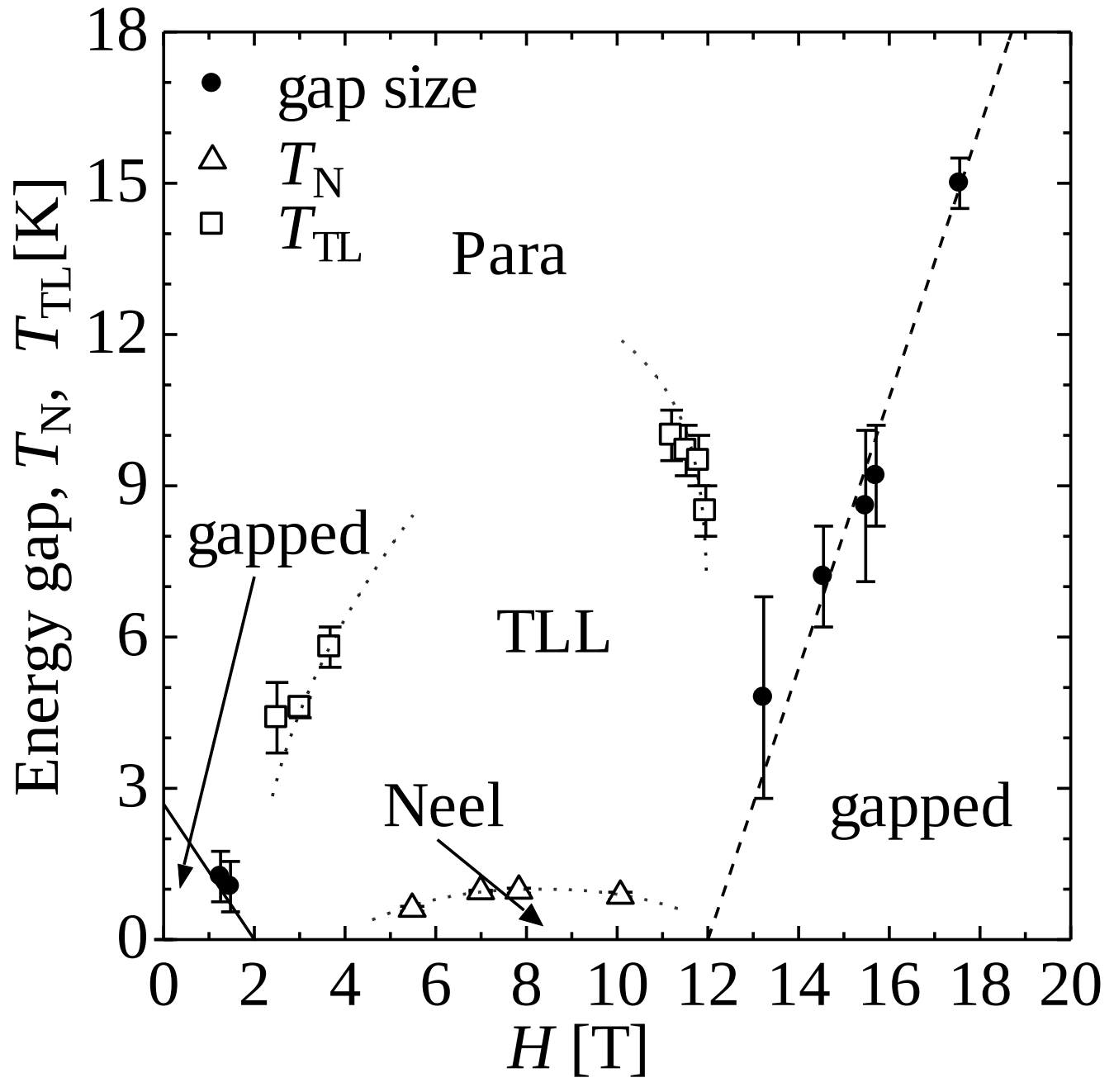

Figure 5 HT diagram for $Rb_2Cu_2Mo_3O_{12}$ showing the gap size under the fields below $H_C \simeq 2$ T and above $H_S \simeq 12$ T, the magnetic ordering temperature $T_N$ below 1 K, and the crossover temperature $T_{TL}$ between the nematic TLL and the paramagnetic state. The solid and dashed lines indicate the $H$-dependence of the gap, $g(H_C - H)\mu_B/k_B$ with $g = 2$ and 4, respectively [19,20]. Dotted curves are eye-guides to identify each phase.